\documentclass[sigconf]{acmart}

\AtBeginDocument{%
  }

\setcopyright{none}
\acmDOI{}
\acmISBN{}

\acmConference[LOCO '24]{1st International Workshop on Low Carbon Computing}{December 1,
  2024}{Glasgow, Scotland, UK}

\begin{document}

\title{On Combining Two Server Control Policies for Energy Efficiency}

\author{Jingze Dai}
\email{david1147062956@gmail.com}
\affiliation{%
  \institution{Department of Computing and Software\\
  McMaster University}
  \city{Hamilton}
  \country{Canada}
}

\author{Douglas G. Down}
\email{downd@mcmaster.ca}
\affiliation{%
  \institution{Department of Computing and Software\\
  McMaster University}
  \city{Hamilton}
  \country{Canada}
}

%
%
%
%
%
%

\renewcommand{\shortauthors}{Dai and Down}

\begin{abstract}
Two popular server control policies are available for reducing energy consumption while maintaining acceptable performance levels: server speed scaling and the ability to turn servers off (and on). In this work, we explore the question of whether there are synergistic effects between these two mechanisms. To do this, we employ a continuous-time Markov chain model where the server can be turned off (and turning the server back on takes some time) and where the speed of the server can take on two values: a nominal operating speed and a reduced operating speed. For a cost function that is linear in the mean response time and server power consumption, we suggest that the mechanisms are not synergistic in that for all system loads, one mechanism is dominant in that if the other mechanism is also employed, there is only a small decrease in cost.
\end{abstract}

%

%

\maketitle

\section{Introduction}

It is well-known that data centre power consumption is a significant contributor to global power consumption. This issue has been discussed in a number of venues (for example \cite{ei20,jon18}). There is no consensus on exactly what power consumption levels will look like, due to the changing landscape of large language models, the use of more efficient hardware, etc., However, there is agreement that the power consumption will be high. The use of guidelines on how to operate servers in data centres is one of the
common practices that data centre operators use to lower energy consumption without
compromising performance constraints. We are concerned with how to operate a single-server where server speed scaling is employed and the server can be turned off (and on). The understanding of the behaviour of a single-server system
is a useful step in understanding the behaviour of more complex, multi-server systems. The two mechanisms available are: (i) server speed scaling, so that operating policies may
vary server speed to reduce power consumption while meeting performance requirements; and (ii) turning a server off during periods of low utilization. In the latter mechanism, average power consumption is reduced, as idle servers still consume a significant amount of power. However, performance (in terms of response times) is adversely affected due to the time that it takes to turn on a server (the setup time).

In isolation, the two mechanisms described above have been extensively studied; in the interests of space it is not possible to give an exhaustive reference list. Representative of the work studying on/off servers is that of Gandhi et al. \cite{gangupharkoz10} and Maccio and Down \cite{macdow15}. For servers with speed scaling, representative work can be found in George and Harrison \cite{geohar01} and Wierman, Andrew and Tang \cite{wieandtan12}. There do not appear to be many works that combine together these two mechanisms, but there are two that are quite relevant to the work presented in this paper. The first, by Lu, Aalto and Lassila \cite{luaallas13} analyzes three models: a baseline model where the server is always on and the server's processing rate is optimized; a server that turns off when idle and has an optimized processing rate when on; and a server that turns off when idle and uses optimized linear speed scaling (the processing rate depends on the number of waiting jobs in a linear manner). For a cost function that is a linear combination of the mean response time and the power consumption, it is shown that when the mean switching time is low, there is significant benefit from the ability to turn the server off and that it is sufficient to use a single processing rate. When the mean switching time increases, the on/off server with a constant processing rate may perform poorly, while benefits can still be accrued from linear speed scaling. The second paper, by Badian-Pessot, Down and Lewis \cite{baddowlew21}, uses a Markov Decision Process approach to show that for a similar cost function, there exists an optimal policy that either (a) never turns the server off and the server's processing rate increases monotonically with the number of waiting jobs, or (b) turns the server off when idle, waits for a threshold number of jobs to be waiting in the queue before turning on, then uses a processing rate that is monotonically increasing with the number of waiting jobs.

One key question of interest is which mechanism is the most effective at reducing costs, the ability to turn the server off, or the ability to adjust the processing rate? Or is there synergy between these mechanisms that make them more effective when combined? We explore these questions using a continuous-time Markov chain model that describes the behaviour of the system, described in the next section. Following that, we provide an outline of our results.

\section{Model}

\begin{figure}[h]
  \centering
  \includegraphics[width=\linewidth]{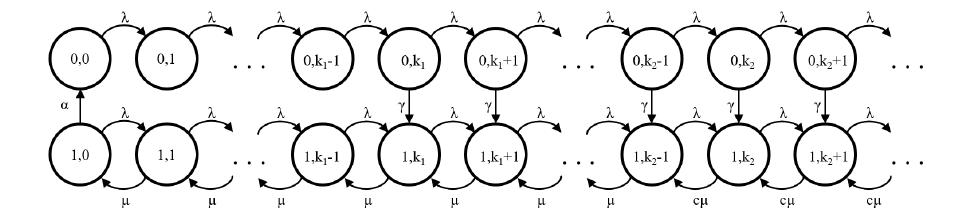}
  \caption{CTMC system model}
  \label{fig:ctmc}
 \end{figure}

We consider a server which can be turned on and off, and when on has two distinct server speeds (processing rates), a fast and a slow speed. This leads to five distinct states for the server. The OFF state represents the server being off;
the switching state represents the server being turned on; the SLOW state means the
server is operating at the slow speed; the FAST state represents the server is operating
at the fast speed; and lastly, the IDLE state represents the server is idle (on with no jobs waiting). We define
additional parameters to indicate state transitions for the server. We define a threshold to turn on the server, $k_1$, where the server will be transferred from OFF to switching
as soon as there are $k_1$ jobs in the system, and the speed scaling threshold $k_2$
is defined to signal the transition to FAST. Furthermore, the time to setup a server is exponentially
distributed with rate $\gamma$, after which the server moves from switching to either SLOW or
FAST. Lastly, we also define a turnoff delay rate, $\alpha$, that captures how
long the server waits when there are no jobs, before transferring the server from IDLE
to OFF (the value of $\alpha$ can be controlled, in particular $\alpha=0$ corresponds to the server never turning off and $\alpha=\infty$ corresponds to the server turning off instantly). There is a buffer of infinite size and jobs are processed in a FCFS manner. Jobs
arrive to the system following a Poisson process with rate $\lambda$. Jobs can only be processed
in either the SLOW or FAST states, where the processing time is exponentially distributed
with rates $\mu$ and $c\mu$ for SLOW and FAST, respectively, and $c>1$ is the scaled speed
factor. We can model the system as a Continuous Time Markov Chain where the state is given by $(s,q)$.  The first entry, $s$, corresponds to the server state: $s=0$ denotes that the server is off (OFF), while $s=1$ denotes that the server is on (switching, IDLE, SLOW, FAST). The second entry, $q$, gives the number of jobs in the system. An example state transition diagram is given in Figure~\ref{fig:ctmc}.

The server power consumption model is as follows. In server state OFF, no power is consumed. In server states IDLE, switching, SLOW and FAST, the power consumptions are $p_i$, $p_w$, $p_s$, $p_f$, respectively. The reduction in power consumption in moving from $p_s$ and $p_f$ is not linear, power consumption is typically modelled to be a second to third-order function of the processing rate. The mean power consumption (averaged over all server states) is denoted by $E[P]$. If we let $E[R]$ be the mean response time (the time that a job spends in the system from arrival to departure), we are interested in cost functions of the form:
\begin{equation}
E[R]+\beta E[P],
\end{equation}
where $\beta>0$ gives the relative weighting between performance (mean response time) and power consumption.

\section{Results}

It is possible to solve for the steady-state distribution of the CTMC described in the previous section. Some general, explicit results (with very cumbersome expressions) are available, which we plan to make available in a full paper. For any fixed set of parameters, it is of course possible to provide a numerical solution for the steady-state distribution. Extensive numerical experiments have been performed to provide our key observations and guide our analytic work. The key observation is that for all arrival rates, the cost reduction that can be obtained from using both mechanisms is small. The dominant policies can also be described in a fairly clean manner. In order to state the following result, it must be the case that the faster processing rate cannot be always preferred over the slower rate (in other words, $p_f/(c\mu) > p_s/\mu$. It appears that there are a series of thresholds $\lambda_k$ (depending on the underlying system parameters) such that:

\begin{itemize}
\item For arrival rates $\lambda > \lambda_1$, only rate $c\mu$ is used and the server never turns off
\item For arrival rates $\lambda_1 > \lambda > \lambda_2$, both rates $\mu$ and $c\mu$ are used and the server never turns off
\item For arrival rates $\lambda_2 > \lambda > \lambda_3$, only rate $\mu$ is used and the server never turns off
\item For arrival rates $\lambda < \lambda_3$, only rate $\mu$ is used and the server is turned on and off
\end{itemize}

In particular, a key observation from the above is that the two mechanisms are not synergistic in the sense that it is never the case that significant cost reductions can be made by using both processing rates and turning the server on and off.

These results have a number of implications for energy-efficient control. They also raise a number of research questions. In terms of implications, the main observation is that which power-savings mechanism is preferred is quite intuitive and at a given system load, one of the mechanisms is dominant. For lower system loads, one should take advantage of the mechanism that turns the server on and off to avoid the power consumption of relatively long idle periods. As the load on the system gets larger, the server idle periods become shorter (using the fast processing rate) and it becomes more effective to compensate for idleness by simply slowing down the server when the number of jobs in the system is small. So, a simple policy would be to use the on-off mechanism (with the slowest processing rate) below a given system utilization and speed scaling otherwise. One issue with this is that it is often difficult in practice to measure the arrival rate to the system, so it would be worthwhile to explore whether effective policies could be designed that do not use arrival rate information. Another key question is to what degree these insights carry over to multiserver systems. Based on our past experience analyzing multiserver systems, we would expect that these insights for a single-server system would indeed carry over to multi-server systems.

\end{document}